\documentclass[preprint,5p,times,twocolumn]{elsarticle}

\usepackage[numbers]{natbib}
\usepackage{amsmath,amssymb,amsfonts}
\usepackage{algorithmic}
\usepackage{graphicx}
\usepackage{textcomp}
\usepackage{xcolor}
\usepackage{makecell}
\usepackage{multirow}
\usepackage{threeparttable}
\usepackage{bbding}
\usepackage{booktabs}
\usepackage{etoolbox}
\usepackage{orcidlink}
\usepackage{array}
\usepackage{subfigure} % 能用，但比较旧
\usepackage{soul}
\usepackage{hyperref}
\usepackage{pifont}

\usepackage{dblfloatfix}

\def\tsc#1{\csdef{#1}{\textsc{\lowercase{#1}}\xspace}}
\tsc{WGM}
\tsc{QE}
\journal{  }

\begin{document}

\let\WriteBookmarks\relax

\begin{frontmatter}

% \title{}
\title{Triple-Phase Sequential Fusion Network with Contrast Enhancement Kinetics for Hepatobiliary-Phase Liver MRI Synthesis}
\author[radio,yuyue]{Qiuli~Wang\fnref{fn1}}
\ead{wangqiuli@tmmu.edu.cn}

\author[yuyue]{Xinhuan~Sun\fnref{fn1}}
\ead{sunxinhuan@jflab.ac.cn}

\author[radio]{Fengxi~Chen\fnref{fn1}}
\ead{chenfx1990@tmmu.edu.cn}

\author[yuyue]{Yongxu~Liu}
\ead{liuyongxu@jflab.ac.cn}

\author[radio]{Jie~Cheng}
\ead{chengjie@tmmu.edu.cn}

\author[radio,lab]{Lin~Chen}
\ead{821341098@qq.com}

\author[radio]{Jiafei~Chen}
\ead{chenjiafei@tmmu.edu.cn}

\author[BigData]{Siwei~Miao}
\ead{miaosiwei@tmmu.edu.cn}
\renewcommand{\thefootnote}{\fnsymbol{footnote}}

\author[swpu]{Xingpeng~Zhang}
\ead{xpzhang@swpu.edu.cn}

\author[dianzikeda]{Yue~Zhang}
\ead{yue_zhang@uestc.edu.cn}

\author[radio]{Xiaoming~Li}
\ead{lxm359261069@tmmu.edu.cn}

\author[radio,yuyue]{Wei~Chen\fnref{cor1}}
\ead{landcw@tmmu.edu.cn}

%% 共一作者脚注
% \fntext[fn1]{These authors contributed equally to this work.}
\fntext[fn1]{Q.~Wang, X.~Sun, and F.~Chen contribute equally to this manuscript.}

%% 通讯作者脚注
\fntext[cor1]{Corresponding authors: Y.~Zhang, X.~Li, and W.~Chen.}
% \cortext[cor2]{Principal corresponding author}
\renewcommand{\thefootnote}{\arabic{footnote}}

%% 作者单位1
\affiliation[radio]{organization={7T Magnetic Resonance Translational Medicine Research Center, Department of Radiology, The First Affiliated Hospital (Southwest Hospital) of Army Medical University},
            % addressline={},
            city={Chongqing},
            postcode={400033}, 
            country={China}}

% %% 作者单位2
\affiliation[yuyue]{organization={Yu-Yue Pathology Research Center, Jinfeng Laboratory},
            % addressline={},
            city={Chongqing},
            postcode={400000},
            country={China}}

\affiliation[lab]{organization={Department of Clinical Laboratory Medicine, The First Affiliated Hospital (Southwest Hospital) of Army Medical University},
            city={Chongqing},
            postcode={400000},
            country={China}}

\affiliation[BigData]{organization={Centre for Medical Big Data and Artificial Intelligence, The First Affiliated Hospital (Southwest Hospital) of Army Medical University},
            city={Chongqing},
            postcode={400000},
            country={China}}
            
\affiliation[swpu]{organization={School of Computer Science and Software Engineering, Southwest Petroleum University},
            % addressline={},
            city={Chengdu},
            postcode={610500},
            country={China}}
% %% 作者单位4
\affiliation[dianzikeda]{organization={Laboratory of Intelligent Collaborative Computing, University of Electronic Science and Technology of China},
            % addressline={},
            city={Chengdu},
            postcode={610000},
            country={China}}

%% Abstract
\begin{abstract}  
  Gadoxetate disodium-enhanced MRI is essential for the detection and characterization of hepatocellular carcinoma. However, acquisition of the hepatobiliary phase (HBP) requires a prolonged post-contrast delay, which reduces workflow efficiency and increases the risk of motion artifacts.
  In this study, we propose a Triple-Phase Sequential Fusion Network (TriPF-Net) with contrast enhancement kinetics to synthesize HBP images by leveraging the sequential information from pre-HBP sequences: while T1-weighted imaging serves as the indispensable baseline, the model adaptively integrates arterial-phase (AP) and venous-phase (VP) features when available. By modeling the tissue-specific contrast uptake and excretion dynamics across these three phases, TriPF-Net ensures robust HBP synthesis even under the stochastic absence of one or both dynamic contrast-enhanced sequences. 
  In addition, clinical variables, including age, sex, total bilirubin, and albumin, are incorporated to enhance physiological consistency. 
  Compared with conventional methods, TriPF-Net achieved superior performance on datasets from two centers. On the first-center dataset, the model achieved an MAE of 10.65, a PSNR of 23.27, and an SSIM of 0.76. On the second-center dataset, the model was trained independently, and the corresponding values were 12.41, 23.11, and 0.78, respectively. Moreover, lesion-level evaluation showed that the synthesized images achieved contrast comparable to that of real HBP images, with no significant difference in CNR (P = 0.2699). Blinded evaluations confirmed synthesized images were visually comparable to real HBP, with TriPF-Net remaining robust to missing input phases. This flexible solution enhances clinical workflow and lesion depiction, and may also serve as an automated quality control reference to flag real HBP acquisitions that warrant closer inspection.

\end{abstract}

%% Keywords

\end{frontmatter}

\section{Introduction}
Hepatocellular carcinoma (HCC) is the sixth most common cancer and the third leading cause of cancer-related death worldwide \cite{bray2024global}. 
Early detection and accurate characterization of focal liver lesions are critical for improving prognosis \cite{han2022declining, kamaya2024li}. In clinical practice, gadoxetate disodium-enhanced MRI (Gd-EOB-DTPA MRI) plays an important role in the detection, characterization, and surgical planning of HCC lesions \cite{yan2023deep}. A complete examination typically includes precontrast T1-weighted imaging, arterial-phase (AP) imaging, venous-phase (VP) imaging, and hepatobiliary-phase (HBP) imaging. Among these phases, HBP provides unique functional information for assessing liver function reserve and improving lesion conspicuity \cite{jiang2023vict2}.

\begin{figure*}[!t]
  \centering
  \includegraphics[width=0.9\textwidth]{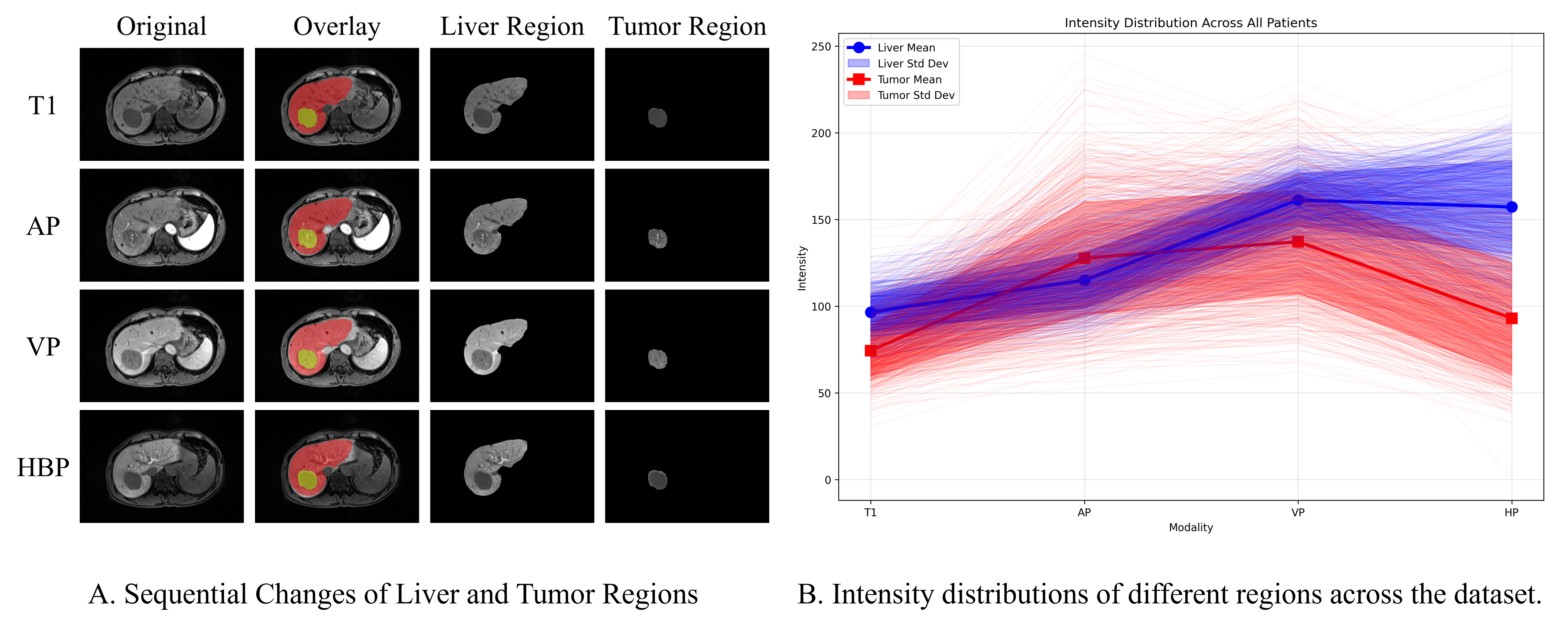}
  % \centerline{\includegraphics[width=0.9\textwidth]{drawbacks.png}}
  \caption{(A) Four-phase Gd-EOB-DTPA MRI (T1, AP, VP, HBP) showing distinct temporal intensity patterns of liver parenchyma (red) and tumor (yellow).
  (B) Mean $\pm$ SD signal intensity across 700 patients (2,800 sequences). Both regions show consistent phase-wise trends, while tumors exhibit faster enhancement and washout than liver parenchyma. These observed kinetics provide the empirical basis for the Contrast Enhancement Kinetics Module described in Section~\ref{sec:cek}.}
  \label{fig1}
  \end{figure*}

Despite its clinical value, HBP imaging acquired with hepatocyte-specific contrast agents requires a prolonged post-contrast delay, which increases patient burden, reduces examination tolerance, and may compromise acquisition consistency in clinical practice. The extended waiting period also increases the risk of motion artifacts and variable image quality. In routine practice, suboptimal HBP acquisitions may go unnoticed until radiologist review, at which point re-acquisition is no longer feasible.

To address these limitations, deep learning-based HBP synthesis has emerged as a promising strategy for generating HBP images from pre-hepatobiliary phases, potentially reducing reliance on delayed acquisition \cite{zhao2020tripartite, haubold2023contrast}. Early studies using the transitional phase (TP) as input in convolutional neural networks or generative adversarial networks (GANs) demonstrated encouraging performance, but their effectiveness depends heavily on the availability and quality of TP images. Subsequent multi-task GAN frameworks extended synthesis to earlier phases and improved downstream tasks such as fibrosis grading, while registration-guided models, such as MrGAN, further emphasized structural consistency \cite{zhao2020tripartite, haubold2023contrast}. More recently, diffusion-based methods have attempted to generate all contrast phases directly from non-contrast MRI, showing promising feasibility but still requiring large datasets and raising concerns about cross-institutional stability.

However, current approaches still face two major limitations: limited flexibility in handling incomplete inputs and insufficient modeling of sequential enhancement dynamics across MRI phases.

(1) Some methods rely solely on a single phase, such as T1-weighted imaging, which limits robustness and generalizability, whereas others require the simultaneous availability of all three pre-HBP phases, namely T1-weighted, arterial-phase (AP), and venous-phase (VP) images, for synthesis. Such rigid input requirements reduce the usability of incomplete clinical datasets and limit the applicability of these models in real-world settings \cite{azad2025addressing}.

(2) Conventional methods often overlook the characteristic enhancement dynamics of liver parenchyma and focal lesions across MRI phases, as illustrated in Fig.~\ref{fig1}. T1-weighted images reflect baseline contrast, AP images often show marked tumor enhancement relative to the liver, and VP images capture progressive parenchymal enhancement while certain tumors exhibit washout or reduced relative intensity. These temporal variations encode physiologically meaningful information about tissue-specific contrast uptake and excretion, which is highly relevant for accurate and clinically consistent HBP synthesis \cite{yoshimitsu2025washout}.
To address these challenges, we propose the Triple-Phase Sequential Fusion Network (TriPF-Net), which leverages a flexible triple-phase input strategy to synthesize high-fidelity HBP images. TriPF-Net consists of three key components: an Enhanced Region-Guided Encoder, a Triple-Phase Unification Module, and a Contrast Enhancement Kinetics Module. The proposed framework offers three main advantages:

(1) It establishes a hierarchical triple-phase configuration. Specifically, T1-weighted imaging serves as the indispensable baseline input, while arterial-phase (AP) and venous-phase (VP) sequences are adaptively integrated as complementary dynamic features when available. This flexible architecture supports robust HBP synthesis under the stochastic absence of one or both contrast-enhanced phases, while maximizing the utility of available vascular-phase information.

(2) It explicitly models sequential enhancement dynamics from pre-HBP phases toward the target HBP image. Unlike static fusion methods, TriPF-Net captures the characteristic signal evolution of liver parenchyma and focal lesions (Fig.~\ref{fig1}A). By learning the temporal transition from pre-contrast T1-weighted imaging to vascular enhancement phases and finally to HBP, the network improves the fidelity of synthesized HBP images, particularly in depicting tissue-specific uptake and washout patterns.

(3) It incorporates multi-dimensional physiological context. By jointly integrating triple-phase imaging data with key clinical variables, including age, sex, and liver function biomarkers such as serum bilirubin and albumin, TriPF-Net generates HBP images that are not only visually realistic but also physiologically plausible, reflecting patient-specific contrast-agent uptake and excretion kinetics.

The contributions of this study are summarized as follows:
\begin{itemize}
  \item We propose TriPF-Net, a flexible framework for HBP MRI synthesis that uses T1-weighted images as the primary input while adaptively incorporating AP and VP information when available. This design enables high-fidelity synthesis under missing-phase conditions and improves robustness in real-world clinical settings.

  \item We introduce an Enhanced Region-Guided Encoder, a Triple-Phase Unification Module, and a Contrast Enhancement Kinetics Module to capture region-specific imaging features, adaptively unify multi-phase information, and model physiologically meaningful enhancement dynamics of liver parenchyma and focal lesions. By further integrating clinical biomarkers such as serum bilirubin and albumin, TriPF-Net improves the physiological plausibility of synthesized HBP images.
  
  \item We validate TriPF-Net on datasets from two clinical centers using quantitative metrics and expert radiologist assessment. The synthesized images achieve lesion-level contrast-to-noise ratios (CNR) comparable to real HBP images ($P = 0.2699$), and blinded visual evaluation further supports their potential to streamline clinical workflows and serve as an automated quality control reference for real HBP acquisitions.

\end{itemize}

\section{Related Work} 
\label{relatedwork}
\subsection{Hepatobiliary-Phase MRI Synthesis}
Hepatobiliary-phase (HBP) MRI acquired with hepatocyte-specific contrast agents (e.g., Gd-EOB-DTPA or gadobenate dimeglumine) is essential for the detection and characterization of hepatocellular carcinoma (HCC) \cite{yu2022gd}. However, HBP acquisition requires a 10--20 min delay after contrast injection, which prolongs examination time, reduces scanner throughput, and increases the risk of motion artifacts \cite{wang2023optimization, taouli2023consensus}. This challenge is particularly pronounced in patients with impaired liver function, such as those with advanced cirrhosis or cholestasis, in whom contrast uptake may be delayed or insufficient \cite{caparroz2022portal}. To alleviate these limitations, several deep learning approaches have been proposed, which are reviewed below.

Existing studies have explored HBP synthesis using regression-based convolutional neural networks, generative adversarial networks (GANs), multi-phase fusion strategies, and attention-based unsupervised learning \cite{zhao2020tripartite, haubold2023contrast}.
Li et al. \cite{li2024image} proposed a multimodal latent diffusion model with a latent spatiotemporal alignment transformer for synthesizing HBP MRI from five early contrast-enhanced phases, achieving high image quality through adaptive cross-modal fusion. 
Liu et al. \cite{liu2023multi} developed MrGAN, a registration-guided GAN that translates multi-phase DCE-MRI into virtual HBP images for liver lesion diagnosis. More recently, Abosabie \cite{abosabie2026deep} introduced a 3D cycle-consistent GAN to synthesize multiphasic contrast-enhanced liver MRI examinations from precontrast T1- and T2-weighted sequences, showing high quantitative and qualitative similarity to ground-truth images.

Overall, these studies have advanced HBP MRI synthesis for HCC and highlighted the value of multi-phase information. However, most methods still lack explicit modeling of sequential enhancement patterns across phases and mainly treat multi-phase inputs as complementary features. In addition, complete and high-quality phase acquisition is often unavailable in clinical practice. These limitations motivate the development of a flexible synthesis framework that can capture sequential dynamics while remaining robust to missing inputs.

\subsection{Deep Generative Models for Medical Image Synthesis}
Deep generative models have become a major technical foundation for medical image synthesis, helping address clinical workflow bottlenecks and data scarcity. Among them, diffusion models have recently attracted considerable attention because of their ability to generate high-fidelity and clinically meaningful images. For example, Pinaya et al. \cite{Pinaya2022BrainIG} introduced a latent diffusion model for brain MRI synthesis that generates anatomically plausible images in latent space. Shi et al. \cite{11194198} further proposed a privacy-preserving latent diffusion framework for realistic synthetic medical image generation, offering a potential solution for secure medical data sharing.

Despite their strong performance, diffusion models remain computationally expensive in both training and inference, which limits their practicality in time-sensitive clinical settings. By contrast, GANs are often more suitable for rapid deployment because of their mature training frameworks, efficient inference, and strong image generation capability \cite{CHEN2022105382}. Yuan et al. \cite{YUAN2020101731} proposed a unified GAN for multimodal segmentation from unpaired 3D medical images, jointly optimizing image synthesis and segmentation. Gao et al. \cite{natmedgao2023synthetic} further demonstrated the value of GAN-generated synthetic data for developing generalizable learning-based algorithms in X-ray image analysis. Although these models have shown promise, challenges remain in ensuring the clinical validity and robustness of synthesized images, especially for multi-phase imaging data.

Nevertheless, most generative models optimize image similarity or adversarial realism, while explicit physiological constraints reflecting contrast enhancement kinetics remain underexplored in liver MRI synthesis.

\subsection{Multi-Phase Fusion and Missing-Modality Learning}
In multi-sequence medical imaging, a single modality often provides limited anatomical and functional information. Consequently, synthesis based on a single input modality may produce images with insufficient fidelity and clinical relevance \cite{Alqutayfi2026MRICS}. To address this issue, many studies have explored multi-phase fusion and missing-modality learning. For example, Li et al. \cite{10.1007/978-3-030-32251-9_87} proposed DiamondGAN, a unified multi-modal GAN capable of synthesizing high-quality target sequences even when some input modalities are missing. Zhou et al. \cite{9004544} developed Hi-Net, a hybrid-fusion network that improves synthesis consistency and fidelity by effectively integrating complementary information from different modalities. Jia et al. \cite{JIA2026103287} further proposed a multi-modal synthesis framework based on dual-branch wavelet encoding and deformable feature interaction to improve performance under missing-modality settings.

To improve robustness to incomplete clinical imaging data, Zhang et al. \cite{Zhang2023UnifiedMI} proposed UNISYN, a unified multi-modal synthesis model that flexibly adapts to missing modalities through optimized multi-phase feature fusion. This model is related to TriPF-Net in that both aim to improve robustness under incomplete inputs. However, Zhang et al. mainly focus on static multi-modal feature fusion and do not explicitly model the sequential dynamics of multi-phase imaging. Different from existing studies, TriPF-Net is designed for Gd-EOB-DTPA-enhanced liver MRI and incorporates a Contrast Enhancement Kinetics Module to explicitly model region-specific enhancement trajectories of liver parenchyma and focal lesions, thereby improving the physiological consistency of synthesized HBP images.

\begin{figure*}[!t]
  \centering
  \includegraphics[width=0.8\textwidth]{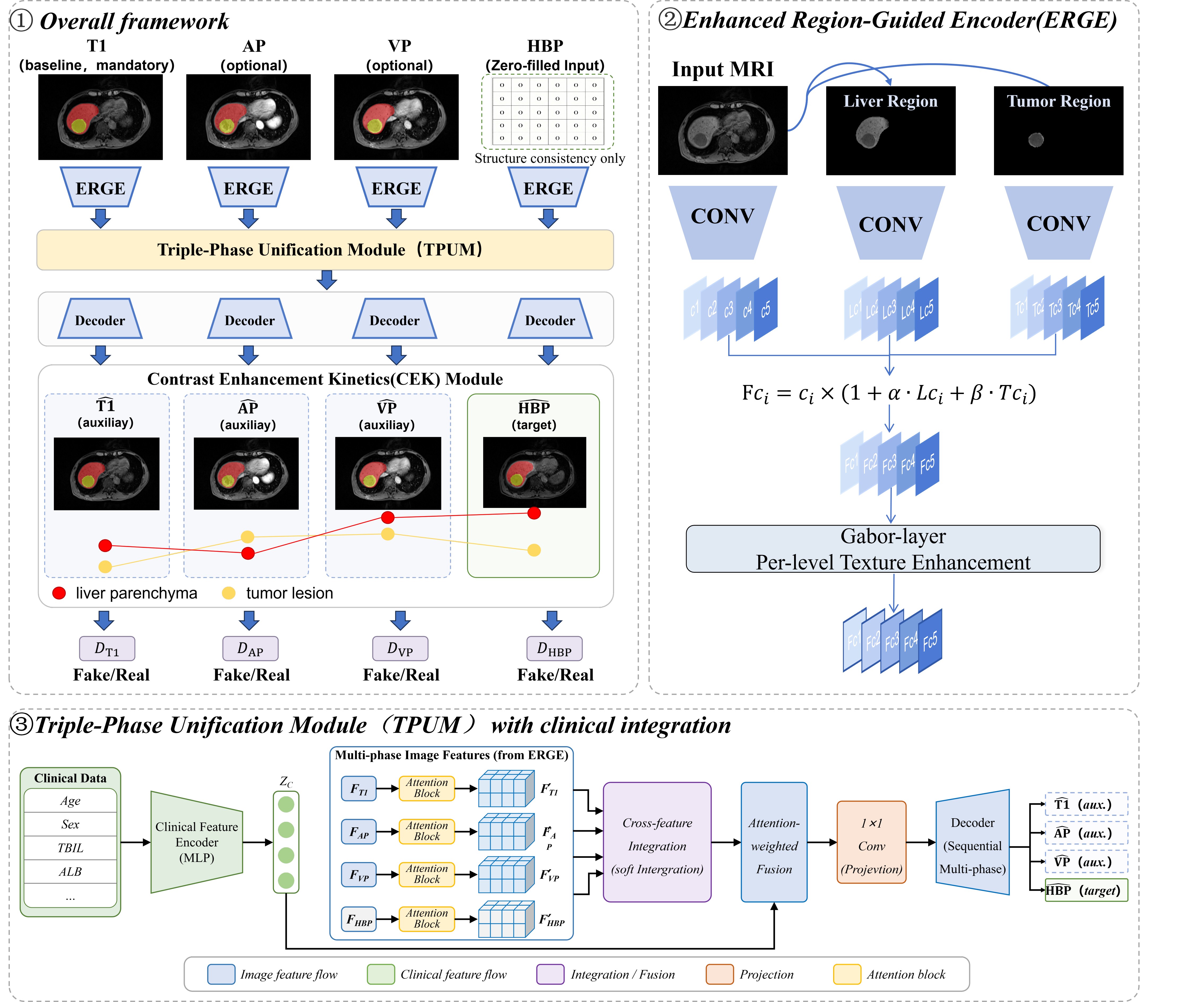}
  % \centerline{\includegraphics[width=0.9\textwidth]{drawbacks.png}}
  \caption{
    Architecture of the proposed Triple-Phase Sequential Fusion Network (TriPF-Net). 
    The framework is organized into three numbered components: 
    \ding{172} overview of the overall synthesis pipeline, in which T1-weighted imaging serves as the mandatory baseline input and AP/VP images are adaptively incorporated when available; 
    \ding{173} Enhanced Region-Guided Encoder, which extracts region-aware features from the liver parenchyma, tumor, and background regions using anatomical masks; 
    and \ding{174} Triple-Phase Unification Module, which adaptively integrates multi-phase imaging features and clinical variables to generate the synthesized HBP image. The four ERGE blocks are independent, isomorphic encoder instances, one per image modality; a phase that is not acquired is fed as a zero-filled input of the corresponding branch, and the HBP branch is therefore always zero-filled. The HBP panel is the ground-truth target used for training-time supervision and is never a network input. The clinical variables are encoded separately by \(\phi(C)\).
    }
   \label{fig2}
  \end{figure*}
  \section{Methods}
  \label{methods}
  
  \subsection{Overview of Triple-Phase Sequential Fusion Network}
  
  The proposed Triple-Phase Sequential Fusion Network (TriPF-Net), as illustrated in Fig.~\ref{fig2}, is designed to synthesize hepatobiliary-phase (HBP) MRI from pre-HBP multi-phase inputs. The framework consists of three major components: an Enhanced Region-Guided Encoder (ERGE), a Triple-Phase Unification Module (TPUM), and a Contrast Enhancement Kinetics Module (CEK Module).
  
  TriPF-Net takes three pre-HBP MRI phases, including T1-weighted, arterial-phase (AP), and venous-phase (VP) images, as inputs. The input set is denoted as:
  \[
  \mathcal{X}_{\mathrm{in}} = \{X_{T_1}, X_{AP}, X_{VP}\}.
  \]
  The HBP image \(X_{HBP}\) is not a network input; it is shown in Fig.~\ref{fig2} only as the training-time target for supervising the synthesized HBP. Accordingly, the encoder comprises one branch per phase of the examination, and a phase that is not acquired is fed as a zero-filled input of the corresponding branch; the HBP branch is therefore always fed with zeros and only acquires information after the phase features have been fused.

  Different from models that only predict the final HBP image, TriPF-Net generates a complete sequential image set:
  \[
  \hat{\mathcal{X}} =
  \{\hat{X}_{T_1}, \hat{X}_{AP}, \hat{X}_{VP}, \hat{X}_{HBP}\},
  \]
  where \(\hat{X}_{HBP}\) is the target synthesized HBP image, while \(\hat{X}_{T_1}\), \(\hat{X}_{AP}\), and \(\hat{X}_{VP}\) are auxiliary phase predictions used to regularize the multi-phase enhancement trajectory. The overall mapping can be formulated as:
  \[
  \hat{\mathcal{X}}
  =
  \mathrm{TriPF\mbox{-}Net}
  (X_{T_1}, X_{AP}, X_{VP}, C),
  \]
  where \(C\) denotes patient-specific clinical variables.
  
  During training, all generated phases are supervised by their corresponding ground-truth images. This multi-phase supervision encourages the network to preserve the sequential relationship among different contrast phases and stabilizes the learning of enhancement kinetics.

  \subsection{Enhanced Region-Guided Encoder}
  
  The Enhanced Region-Guided Encoder (ERGE) is designed to extract region-aware representations from multi-phase liver MRI. It captures complementary features from three anatomical domains: focal lesions, liver parenchyma, and the global abdominal background.
  
  Liver masks (\(M_L\)) are automatically generated using MRSegmentator, while tumor masks (\(M_T\)) are obtained using MedSAM with semi-automatic refinement. Based on these masks, three anatomical region masks are defined:
  \[
  M_{\mathrm{liver}} = M_L \setminus M_T,\quad
  M_{\mathrm{tumor}} = M_T,\quad
  M_{\mathrm{back}} = 1 - M_L.
  \]
  
  For each acquired phase \(p \in \{T_1, AP, VP\}\), ERGE extracts region-guided features as:
  \[
  F_p = E_{\mathrm{ERGE}}(X_p, M_L, M_T),
  \]
  where \(E_{\mathrm{ERGE}}(\cdot)\) denotes the region-guided encoder.
  
  Within ERGE, high-dimensional representative features are extracted using a pretrained ResNet-101 backbone and are further enhanced by multi-directional Gabor filtering. The Gabor filtering layer is introduced to strengthen the representation of local texture, edge information, and fine structural variations, which are important for preserving lesion boundaries and intrahepatic anatomical details. This design allows the encoder to jointly capture local lesion-specific features and global anatomical context. Rather than sharing a single encoder across phases, TriPF-Net instantiates one ERGE per image modality, which is why four separate ERGE blocks are drawn in Fig.~\ref{fig2}. The four instances are isomorphic---they share exactly the same architecture (ResNet-101 backbone, region-guided feature extraction, and Gabor filtering layer)---whereas their parameters are learned independently. This branch-wise design is deliberate rather than a duplication artifact: it prevents the different phases from being compressed into a single shared representation before fusion, so that each acquired modality is learned in a targeted manner, with the baseline contrast of T1, the arterial hyperenhancement of AP, and the progressive parenchymal enhancement and washout of VP each represented by a dedicated encoder. The HBP branch receives a zero-filled input and therefore carries no phase-specific information of its own; it becomes informative only after the phase features are unified in the Triple-Phase Unification Module. Furthermore, we also tested ResViT~\cite{dalmaz2022resvit} but observed inferior performance, which is discussed in Table~\ref{tab:xnyy}.

  \subsection{Triple-Phase Unification Module}
  
  The Triple-Phase Unification Module (TPUM) adaptively integrates encoded features from T1-weighted, arterial-phase, and venous-phase images. It is designed according to a hierarchical input configuration in which T1-weighted imaging provides baseline anatomical information, whereas AP and VP images contribute complementary vascular enhancement features.
  
  The encoded multi-phase features are unified as:
  \[
  F_{\mathrm{TPU}} = \mathcal{U}(F_{T_1}, F_{AP}, F_{VP}, F_{HBP}),
  \]
  where \(\mathcal{U}(\cdot)\) denotes the adaptive triple-phase unification operation. By integrating information from all three pre-HBP phases, TPUM captures both baseline anatomical structure and dynamic enhancement patterns before HBP acquisition.
  
  To further incorporate patient-specific physiological context, clinical variables are introduced into the feature fusion process. Let \(C \in \mathbb{R}^{22}\) denote the clinical feature vector, which includes demographic variables (age, sex), liver function and coagulation indicators (e.g., total bilirubin, albumin, ALT, AST, ALP, PT, APTT), platelet count, derived scores (ALBI, FIB-4, APRI), hepatitis B biomarkers (HBsAg, HBV-DNA), interaction terms, and binned total bilirubin categories. These are encoded by a learnable clinical feature encoder:
  \[
  F_C = \phi(C),
  \]
  where \(\phi(\cdot)\) maps the clinical vector into spatial feature representations compatible with imaging features. Specifically, a multi-layer perceptron encodes \(C\) into channel-wise affine coefficients, and the resulting modulation feature maps are added to the imaging features with a small residual weight of 0.1, so that the patient-level physiological condition modulates the overall signal intensity without diluting the image features at the channel level. These variables are used as patient-level physiological context rather than as predictors of specific demographic groups, and age and sex are standard demographic descriptors (age is also a component of the FIB-4 score). Their observed distribution nevertheless reflects the two-center Chinese cohort studied here, and validation on external populations is warranted before broader deployment.

  The final unified representation is obtained by integrating imaging features and clinical features:
  \[
  F_{\mathrm{final}} = \mathcal{G}(F_{\mathrm{TPU}}, F_C),
  \]
  where \(\mathcal{G}(\cdot)\) denotes the multi-scale integration operation. The decoder then generates the complete multi-phase image sequence:
  \[
  \hat{\mathcal{X}}
  =
  D(F_{\mathrm{final}})
  =
  \{\hat{X}_{T_1}, \hat{X}_{AP}, \hat{X}_{VP}, \hat{X}_{HBP}\},
  \]
  where \(D(\cdot)\) denotes the synthesis decoder. The final output used for HBP replacement or clinical evaluation is \(\hat{X}_{HBP}\).

  \subsection{Contrast Enhancement Kinetics Module}
  \label{sec:cek}
  \begin{figure}[!t]
    \centering
    \includegraphics[width=0.4\textwidth]{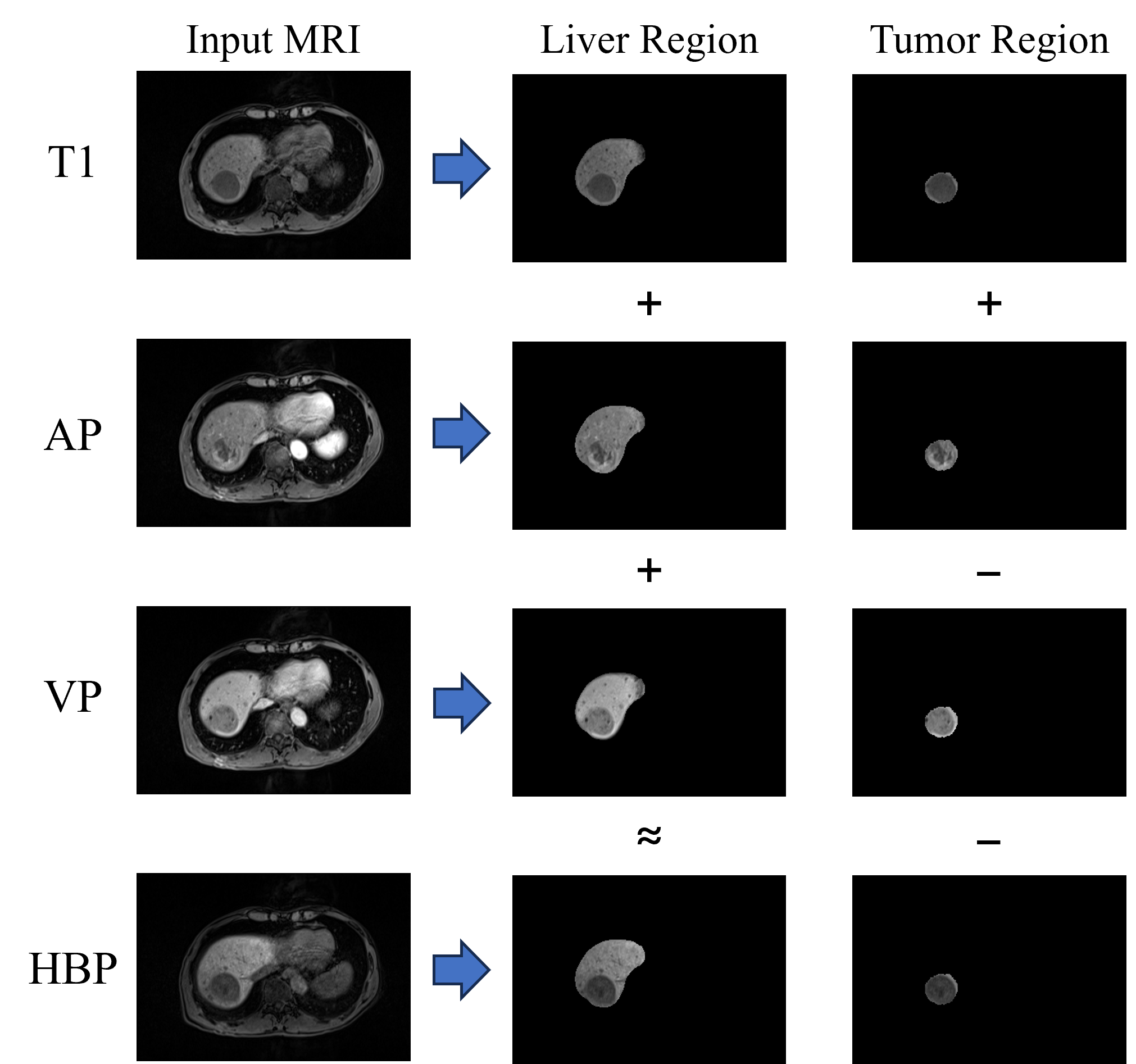}
    \caption{Illustration of the proposed Contrast Enhancement Kinetics Module. For each MRI phase, liver parenchyma and tumor regions are extracted to construct region-specific contrast-enhancement trajectories. The module models the characteristic rapid wash-in and wash-out pattern of HCC lesions and constrains both intra-region temporal trends and inter-region relative intensity relationships. The plus and minus signs indicate expected signal increase and decrease between adjacent phases, respectively, guiding the synthesized multi-phase images toward physiologically plausible enhancement kinetics.}
    \label{fig3}
  \end{figure}
  
  To improve the physiological consistency of synthesized HBP images, we introduce a Contrast Enhancement Kinetics Module (CEK Module). As illustrated in Fig.~\ref{fig3}, liver parenchyma and HCC lesions exhibit distinct signal-enhancement patterns after contrast injection. Motivated by this observation, the CEK Module is designed to learn these region-specific enhancement kinetics and operates on the generated multi-phase image sequence rather than constraining only the final HBP image: The statistics in Fig.~\ref{fig1}B (700 patients, 2,800 phase sequences) show that tumors enhance and wash out faster than the surrounding parenchyma, and the CEK Module is designed to encode exactly these region-specific dynamics.
  \[
  \hat{\mathcal{X}} =
  \{\hat{X}_{T_1}, \hat{X}_{AP}, \hat{X}_{VP}, \hat{X}_{HBP}\}.
  \]
  This design enables the model to learn a sequential enhancement trajectory from pre-contrast T1-weighted imaging to vascular enhancement phases and finally to the hepatobiliary phase.
  
  The CEK Module explicitly models region-specific contrast-enhancement trajectories across MRI phases and imposes clinically motivated constraints on signal-intensity evolution. Specifically, it constrains both intra-region temporal trends and inter-region relative intensity relationships between liver parenchyma and tumor regions.
  
  We define two foreground anatomical regions for kinetics modeling: liver parenchyma (\(R_L\)) and tumor (\(R_T\)):
  \[
  R_L = M_L \setminus M_T, \quad R_T = M_T,
  \]
  where \(M_L\) and \(M_T\) denote the liver and tumor masks, respectively.
  
  For each region \(r \in \{L,T\}\) and phase \(p \in \{T_1, AP, VP, HBP\}\), the regional mean intensity of the generated image is computed as:
  \[
  \hat{s}_r^p =
  \frac{1}{|R_r|}
  \sum_{i \in R_r}
  \hat{X}_p(i).
  \]
  
  The regional mean intensity of the corresponding real image is computed as:
  \[
  s_r^p =
  \frac{1}{|R_r|}
  \sum_{i \in R_r}
  X_p(i).
  \]
  
  The generated region-specific contrast-enhancement trajectory is defined as:
  \[
  \hat{S}_r =
  [\hat{s}_r^{T_1}, \hat{s}_r^{AP}, \hat{s}_r^{VP}, \hat{s}_r^{HBP}].
  \]
  The real trajectory is defined as:
  \[
  S_r =
  [s_r^{T_1}, s_r^{AP}, s_r^{VP}, s_r^{HBP}].
  \]

  \subsubsection{Tumor Temporal Trend Constraint}
  
  In Gd-EOB-DTPA-enhanced liver MRI, HCC lesions often exhibit arterial phase hyperenhancement followed by washout or relative hypointensity in later phases. Therefore, the generated tumor trajectory is encouraged to follow:
  \[
  \hat{s}_T^{AP} \gtrsim \hat{s}_T^{T_1}, \quad
  \hat{s}_T^{AP} \gtrsim \hat{s}_T^{VP}, \quad
  \hat{s}_T^{VP} \gtrsim \hat{s}_T^{HBP}.
  \]
  
  The corresponding tumor temporal trend loss is formulated using hinge-style one-sided penalties:
  \[
  \begin{aligned}
  L_{\mathrm{trend,tumor}}
  =&\
  \max(0, \hat{s}_T^{T_1} - \hat{s}_T^{AP}) \\
  &+
  \max(0, \hat{s}_T^{VP} - \hat{s}_T^{AP}) \\
  &+
  \max(0, \hat{s}_T^{HBP} - \hat{s}_T^{VP}).
  \end{aligned}
  \]
  The first term encourages arterial enhancement relative to the pre-contrast phase. The second term penalizes violation of the expected decrease from AP to VP. The third term encourages further reduction or relative hypointensity from VP to HBP.

  \subsubsection{Liver Parenchyma Temporal Trend Constraint}
  
  In the hepatobiliary phase, liver parenchymal signal may continue to increase in patients with preserved hepatocyte function, but in patients with impaired liver function—common in the HCC population—HBP parenchymal enhancement often remains comparable to that of the venous phase. Accordingly, the model adopts a conservative constraint:
    \[
  \hat{s}_L^{VP} \gtrsim \hat{s}_L^{T_1}, \quad
  \hat{s}_L^{HBP} \approx \hat{s}_L^{VP}.
  \]
  
  The liver temporal trend loss is defined as:
  \[
  \begin{aligned}
  L_{\mathrm{trend,liver}}
  =&\
  \max(0, \hat{s}_L^{T_1} - \hat{s}_L^{VP}) \\
  &+
  \max\left(0,
  \left|
  \hat{s}_L^{HBP} - \hat{s}_L^{VP}
  \right|
  - \delta_L
  \right),
  \end{aligned}
  \]
  where \(\delta_L\) is a tolerance margin that allows small physiological intensity differences between VP and HBP. In this study, \(\delta_L\) was empirically set to 0.05 in the normalized intensity space.
  The first term encourages enhancement from the pre-contrast phase to the venous phase, while the second term encourages the synthesized HBP liver parenchymal intensity to remain close to the generated VP intensity without enforcing strict equality.

  \subsubsection{Liver-Tumor Relative Intensity Constraint}
  
  In addition to intra-region temporal trends, the CEK Module constrains the relative intensity relationship between liver parenchyma and tumor regions. Clinically, HCC lesions may appear hyperintense relative to liver parenchyma in the arterial phase, whereas they often become relatively hypointense in the hepatobiliary phase. Therefore, the generated images are encouraged to satisfy:
  \[
  \hat{s}_T^{AP} \gtrsim \hat{s}_L^{AP}, \quad
  \hat{s}_L^{HBP} \gtrsim \hat{s}_T^{HBP}.
  \]
  
  The corresponding relative contrast loss is defined as:
  \[
  L_{\mathrm{relative}}
  =
  \max(0, \hat{s}_L^{AP} - \hat{s}_T^{AP})
  +
  \max(0, \hat{s}_T^{HBP} - \hat{s}_L^{HBP}).
  \]

  \subsubsection{Overall Loss Functions}
  \label{sec:loss}
  
  To directly supervise the generated images, we define a multi-phase reconstruction loss:
  \[
  L_{\mathrm{rec}}
  =
  \sum_{p \in \{T_1, AP, VP, HBP\}}
  \left\|
  \hat{X}_p - X_p
  \right\|_1.
  \]
  Although the final clinical target is \(\hat{X}_{HBP}\), auxiliary supervision on pre-HBP phases encourages the decoder to preserve the sequential relationship among different contrast phases and stabilizes the learning of enhancement kinetics.

  In addition to reconstruction and contrast-enhancement kinetics constraints, we introduce an adversarial learning strategy to improve the visual realism of the generated images. One discriminator \(D_p\) is instantiated for each phase \(p \in \{T_1, AP, VP, HBP\}\), receiving either the real image \(X_p\) or the generated image \(\hat{X}_p\). The four discriminators are isomorphic instances that share the same architecture but are optimized independently, and they are used only during training and discarded at inference. The adversarial loss for the discriminators is defined as:
  \[
  \begin{aligned}
  L_D =
  -\sum_{p \in \{T_1, AP, VP, HBP\}}
  \Big(
  &\mathbb{E}_{X_p}
  \left[
  \log D_p(X_p)
  \right] \\
  &+
  \mathbb{E}_{\hat{X}_p}
  \left[
  \log \left(1 - D_p(\hat{X}_p)\right)
  \right]
  \Big).
  \end{aligned}
  \]
  
  The adversarial loss for the generator is defined as:
  \[
  L_{\mathrm{adv}}
  =
  -\sum_{p \in \{T_1, AP, VP, HBP\}}
  \mathbb{E}_{\hat{X}_p}
  \left[
  \log D_p(\hat{X}_p)
  \right].
  \]
  
  The final CEK loss is formulated as:
  \[
  L_{\mathrm{CEK}}
  =
  L_{\mathrm{trend,tumor}}
  +
  L_{\mathrm{trend,liver}}
  +
  L_{\mathrm{relative}}.
  \]
  
  The overall training objective for TriPF-Net is:
  \[
  L_{\mathrm{total}}
  =
  L_{\mathrm{rec}}
  +
  \lambda_{\mathrm{CEK}} L_{\mathrm{CEK}}
  +
  \lambda_{\mathrm{adv}} L_{\mathrm{adv}},
  \]
  where \(\lambda_{\mathrm{CEK}}\) and \(\lambda_{\mathrm{adv}}\) control the contributions of the contrast enhancement kinetics constraint and the adversarial loss, respectively. Based on empirical performance, \(\lambda_{\mathrm{CEK}}\) and \(\lambda_{\mathrm{adv}}\) were manually set to 50 and 0.8, respectively.
  
  During optimization, the generator and the discriminators are trained alternately. The discriminators are optimized by minimizing \(L_D\), whereas the generator is optimized by minimizing \(L_{\mathrm{total}}\). By combining multi-phase reconstruction, contrast enhancement kinetics constraints, and adversarial learning, TriPF-Net is encouraged to generate multi-phase images that are anatomically faithful, physiologically plausible, and visually realistic.

\section{Experiments and Results}

\subsection{Study Population}

This retrospective study was approved by the Institutional Review Board of Southwest Hospital, and the requirement for informed consent was waived. Two datasets of patients with HCC who underwent Gd-EOB-DTPA-enhanced MRI were retrospectively collected from the PACS databases of two tertiary hospitals. Dataset A comprised patients enrolled between January 2021 and March 2025, while Dataset B included patients from September 2016 to November 2018.

The inclusion criteria were as follows:  
(1) pathologically confirmed HCC after surgery or clinically diagnosed HCC according to the AASLD/EASL guidelines;  
(2) complete Gd-EOB-DTPA-enhanced MRI examination;  
(3) availability of complete clinical information, including demographic variables and liver function biomarkers.

The exclusion criteria were as follows:  
(1) severe motion or metal artifacts rendering the images non-diagnostic;  
(2) prior locoregional treatment, such as TACE or RFA, that significantly altered hepatic perfusion;  
(3) diffuse infiltrative HCC precluding reliable lesion boundary delineation;  
(4) absence of key laboratory indicators within 1 week before or after MRI.

Finally, Dataset A (790 patients) and Dataset B (507 patients) were independently divided into training, validation, and test sets at a ratio of 7:2:1.

\subsection{MRI Acquisition and Data Collection}

\begin{table}[!t]
  \centering
  \caption{Summary of the two-center datasets used in this study. }
  \label{tab:dataset_summary}
  \begin{tabular}{lcccc}
  \hline
  \textbf{Center} & \textbf{Cases} & \textbf{Train} & \textbf{Val} & \textbf{Test} \\
  \hline
  Dataset A & 790 & 553 & 158 & 79 \\
  Dataset B & 507 & 355 & 101 & 51 \\
  \hline
  \end{tabular}
\end{table}

As shown in Table~\ref{tab:dataset_summary}, image acquisition was performed at two institutions using multiple scanner platforms to ensure data diversity. At the First Affiliated Hospital of Army Medical University (Dataset A), all examinations were performed on a 3.0-T scanner (MAGNETOM Trio Tim, Siemens Healthcare). At the First Affiliated Hospital of Nanjing Medical University (Dataset B), examinations were performed on both 3.0-T scanners (Skyra, Siemens Healthcare; Discovery MR750w, GE Healthcare) and 1.5-T scanners (HDxt Signa, GE Healthcare; uMR 570, United Imaging Healthcare). Each case contains a complete Gd-EOB-DTPA-enhanced MRI examination, including T1-weighted, arterial-phase, venous-phase, and hepatobiliary-phase sequences.

A standardized multiparametric protocol was implemented across scanners, including T2-weighted imaging, dual-echo T1-weighted imaging (in-phase and opposed-phase), and T1-weighted three-dimensional gradient-echo sequences (VIBE for Siemens systems and LAVA for GE systems). Contrast enhancement was performed by intravenous injection of Gd-EOB-DTPA (Primovist, Bayer Schering Pharma) at a standard dose of 0.025 mmol/kg and a flow rate of 1.0 mL/s, followed by a 30-mL saline flush.

Dynamic multiphase imaging was acquired using T1-weighted 3D-VIBE or LAVA sequences at the following time points: arterial phase (AP), initiated by bolus tracking upon contrast arrival in the abdominal aorta; venous phase (VP; portal venous phase), acquired at 55--70 s; transitional phase (TP), acquired at 150--180 s; and hepatobiliary phase (HBP), acquired 15 min after contrast injection.

\subsection{Evaluation Metrics}

Objective evaluation was performed using both global image-level and lesion-level metrics. MAE, PSNR, SSIM, and FID were used to quantify reconstruction fidelity, structural similarity, and distributional consistency between synthesized and ground-truth HBP images.

For lesion-level evaluation, we further calculated the contrast-to-noise ratio (CNR) and signal-to-noise ratio (SNR) to assess lesion conspicuity and tumor-region signal characteristics. The analysis was performed using the slice with the largest tumor area and the two adjacent slices on each side, yielding five slices per patient. Higher CNR indicates better contrast between the lesion and surrounding liver parenchyma, whereas higher SNR indicates more favorable tumor-region signal quality.

\subsection{Subjective Clinical Validation}

Blinded subjective evaluation was conducted by two senior radiologists, each with more than 10 years of experience in abdominal imaging. The readers independently reviewed synthesized and real HBP images and assessed image quality, anatomical detail preservation, and lesion detectability. Discrepancies were resolved by consensus with a third senior radiologist with more than 20 years of clinical experience.

\subsection{HBP Synthesis Performance}

For each dataset, all models were trained, validated, and tested independently using the corresponding center-specific split.
To evaluate the effectiveness of TriPF-Net, we compared it with several representative image synthesis methods using MAE, PSNR, SSIM, and FID. TriPF-Net synthesized HBP images from T1, AP, and VP sequences, whereas TriPF-Net (T1-only) used only T1-weighted images as input. In the one-to-one synthesis setting, Pix2Pix~\cite{Isola_2017_CVPR}, UNISYN~\cite{Zhang2023UnifiedMI} (T1-only), and Li-DiffNet~\cite{Zhang2026LiDiffNet} were included as baselines. In the many-to-one setting, UNISYN~\cite{Zhang2023UnifiedMI} was used for comparison.

\begin{table*}[!t]
\centering
\caption{Quantitative comparison of TriPF-Net and competing methods on Datasets A and B.}
\label{tab:xnyy}
\begin{tabular}{lccccc}
\hline
\textbf{Dataset} & \textbf{Models} & \textbf{MAE $\downarrow$} & \textbf{PSNR $\uparrow$} & \textbf{SSIM $\uparrow$} & \textbf{FID $\downarrow$} \\
\hline
\multirow{7}{*}{Dataset A}
& Pix2Pix~\cite{Isola_2017_CVPR} & $12.49 \pm 4.44$ & $21.99 \pm 2.18$ & $0.716 \pm 0.089$ & 13.45 \\
& ResViT~\cite{dalmaz2022resvit} & $12.82 \pm 4.99$ & $22.00 \pm 2.26$ & $0.690 \pm  0.083$ &  31.03 \\
& UNISYN (T1-only)~\cite{Zhang2023UnifiedMI} & $11.88 \pm 4.19$ & $22.34 \pm 2.18$ & $0.734 \pm 0.087$ & 12.05 \\
& Li-DiffNet~\cite{Zhang2026LiDiffNet} & $10.78 \pm 4.10$ & $23.26 \pm 2.36$ & $0.747 \pm 0.080$ & 12.56 \\
& UNISYN~\cite{Zhang2023UnifiedMI} & $11.31 \pm 4.02$ & $22.82 \pm 2.19$ & $0.747 \pm 0.083$ & 11.32 \\
& TriPF-Net (T1-only) & $11.29 \pm 4.09$ & $22.76 \pm 2.18$ & $0.739 \pm 0.082$ & 10.74 \\
& TriPF-Net & $\mathbf{10.65 \pm 3.84}$ & $\mathbf{23.27 \pm 2.22}$ & $\mathbf{0.761 \pm 0.077}$ & \textbf{10.49} \\
\hline
\multirow{7}{*}{Dataset B}
& Pix2Pix~\cite{Isola_2017_CVPR} & $14.45 \pm 7.33$ & $21.60 \pm 3.38$ & $0.742 \pm 0.125$ & 22.36 \\
& ResViT~\cite{dalmaz2022resvit} & $16.46 \pm 7.21$ & $20.48 \pm 3.00 $ & $ 0.692 \pm 0.114$ &  52.04 \\
& UNISYN (T1-only)~\cite{Zhang2023UnifiedMI} & $14.54 \pm 7.33$ & $21.46 \pm 3.37$ & $0.728 \pm 0.126$ & 23.17 \\
& Li-DiffNet~\cite{Zhang2026LiDiffNet} & $18.16 \pm 12.14$ & $20.67 \pm 4.23$ & $0.647 \pm 0.215$ & 65.18 \\
& UNISYN~\cite{Zhang2023UnifiedMI} & $12.73 \pm 7.13$ & $22.80 \pm 3.74$ & $0.769 \pm 0.120$ & 20.72 \\
& TriPF-Net (T1-only) & $13.34 \pm 7.05$ & $22.24 \pm 3.57$ & $0.754 \pm 0.123$ & 20.54 \\
& TriPF-Net & $\mathbf{12.41 \pm 7.27}$ & $\mathbf{23.11 \pm 3.82}$ & $\mathbf{0.783 \pm 0.120}$ & \textbf{18.71} \\
\hline
\end{tabular}
\end{table*}

\begin{figure*}[!t]
  \centering
  \includegraphics[width=1\textwidth]{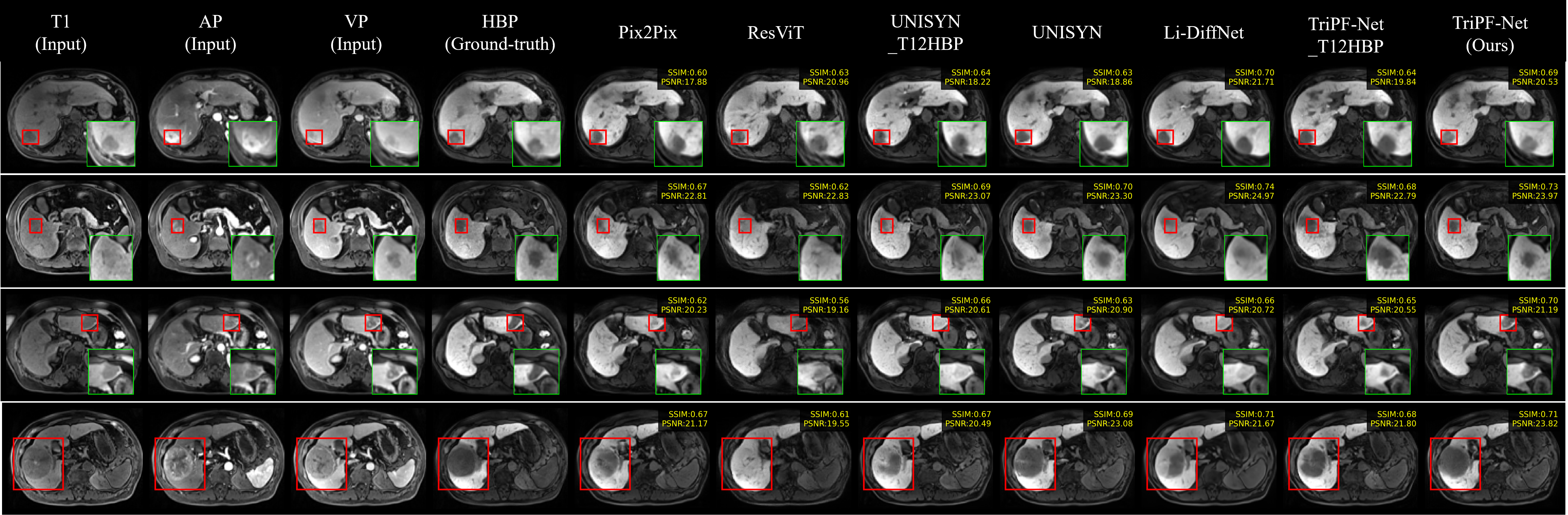}
  \caption{Visual examples of synthetic images generated by TriPF-Net and competing methods on Dataset A. Red boxes highlight the tumor regions, and green boxes provide enlarged views of the corresponding areas to facilitate visual assessment and comparison.}
  \label{fig4}
  
\end{figure*}

\begin{figure*}[!t]
  \centering
  \includegraphics[width=1\textwidth]{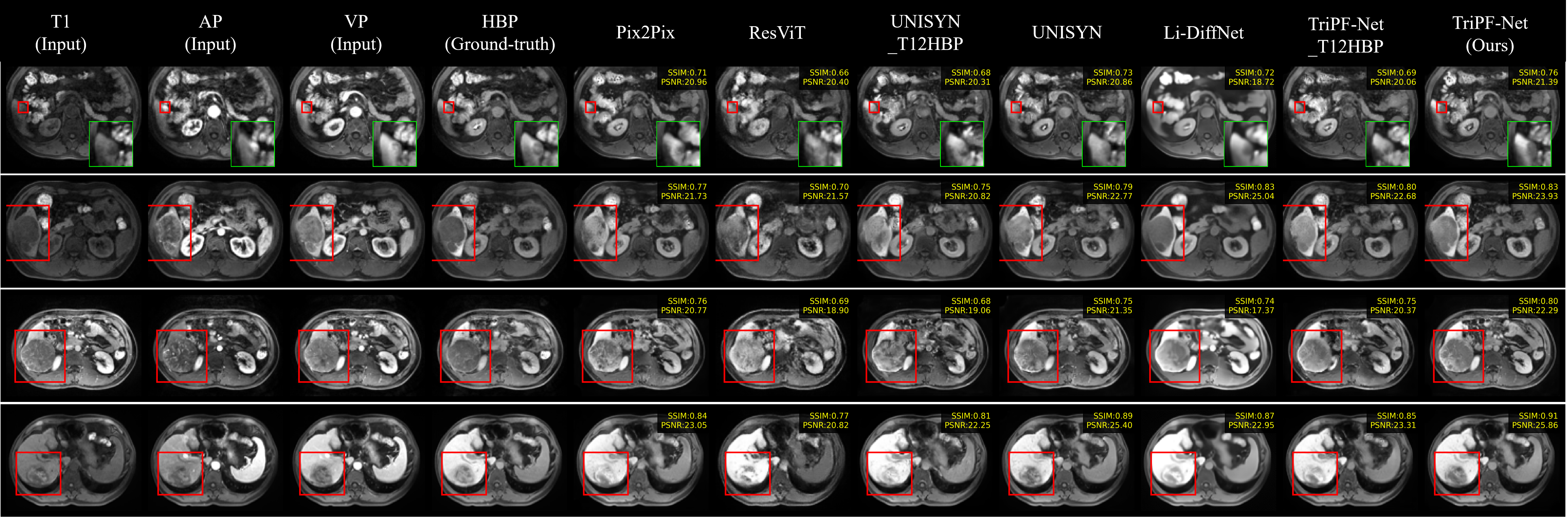}
  \caption{Visual examples of synthetic images generated by TriPF-Net and competing methods on Dataset B. Red boxes highlight the tumor regions, and green boxes provide enlarged views of the corresponding areas to facilitate visual assessment and comparison.}
  \label{fig5}
\end{figure*}

As shown in Table~\ref{tab:xnyy}, TriPF-Net achieved the best overall performance across both datasets, consistently outperforming all baselines in MAE, PSNR, SSIM, and FID. These results indicate that the proposed framework improves reconstruction fidelity, structural similarity, and distributional consistency compared with the competing methods. Fig.~\ref{fig4} and Fig.~\ref{fig5} present visual examples of synthetic images generated by TriPF-Net and competing methods on Dataset A and Dataset B, respectively.

Of note, ResViT~\cite{dalmaz2022resvit} led to a marked drop in SSIM and an increase in FID (e.g., FID increased from 13.45 to 31.03 on Dataset A and from 22.36 to 52.04 on Dataset B; Table~\ref{tab:xnyy}). We therefore retained the CNN-based UNISYN~\cite{Zhang2023UnifiedMI} architecture as the backbone of TriPF-Net.

Notably, within TriPF-Net, multi-sequence input consistently outperformed T1-only synthesis, confirming the importance of incorporating AP and VP information for HBP generation. Although TriPF-Net (T1-only) achieved competitive FID values, its performance on the other metrics remained inferior to that of the full TriPF-Net, indicating that complementary vascular-phase information improves synthesis quality.

\subsection{Robustness to Missing Input Phases}

Because AP and VP acquisitions are not always available in routine practice, we further evaluated TriPF-Net under all combinations of available input phases. In addition to the T1-only and the full three-phase configurations already reported in Table~\ref{tab:xnyy}, two additional models were trained using T1+AP and T1+VP as inputs, so that the two intermediate configurations are also covered. All experiments in this subsection were performed on Dataset A.

\begin{table*}[!t]
  \centering
  \caption{Effect of the available input phase combination on HBP synthesis performance (Dataset A).}
  \label{tab:missing_phase}
  \setlength{\tabcolsep}{4pt}
  \begin{tabular}{lcccc}
  \hline
  \textbf{Input phases} & \textbf{MAE $\downarrow$} & \textbf{PSNR $\uparrow$} & \textbf{SSIM $\uparrow$} & \textbf{FID $\downarrow$} \\
  \hline
  T1 & $11.29 \pm 4.09$ & $22.76 \pm 2.18$ & $0.739 \pm 0.082$ & 10.74 \\
  T1+AP & $11.07 \pm 4.09$ & $22.99 \pm 2.19$ & $0.754 \pm 0.077$ & 10.75 \\
  T1+VP & $10.95 \pm 3.98$ & $23.07 \pm 2.20$ & $0.757 \pm 0.075$ & 10.48 \\
  T1+AP+VP & $\mathbf{10.65 \pm 3.84}$ & $\mathbf{23.27 \pm 2.22}$ & $\mathbf{0.761 \pm 0.077}$ & \textbf{10.49} \\
  \hline
  \end{tabular}
\end{table*}

As shown in Table~\ref{tab:missing_phase}, MAE, PSNR, and SSIM improved monotonically with the number of available input phases. Adding AP or VP alone to the T1 baseline reduced MAE from 11.29 to 11.07 and 10.95, respectively, and using all three phases yielded the best performance (MAE 10.65, PSNR 23.27, SSIM 0.761). The two intermediate configurations performed similarly to each other, indicating that AP and VP provide complementary but partly interchangeable enhancement information. FID, in contrast, showed no clear trend across the four configurations, suggesting that distributional consistency is governed mainly by the synthesis architecture rather than by the number of input phases. Importantly, none of the configurations collapsed: even with T1-weighted imaging as the only input, the loss relative to the full-input setting remained limited (MAE 11.29 vs.\ 10.65, PSNR 22.76 vs.\ 23.27). These results confirm that TriPF-Net degrades gracefully and operates reliably under arbitrary phase availability, which is the clinical scenario it was designed for.

\subsection{Lesion-Level Quantitative Analysis}

To further evaluate lesion conspicuity and tumor-region synthesis quality, we performed lesion-level analysis using CNR and SNR.

\begin{table*}[!t]
\centering
\caption{Lesion-level comparison of CNR and SNR on Datasets A and B.}
\label{tab:cnr_snr}
\begin{tabular}{llcc}
\hline
\textbf{Dataset} & \textbf{Models} & \textbf{CNR (mean $\pm$ std, $P$ value)} & \textbf{SNR (mean $\pm$ std)} \\
\hline
\multirow{7}{*}{Dataset A}
& Pix2Pix~\cite{Isola_2017_CVPR} & $0.94 \pm 0.57$ ($P = 0.0113$) & $-0.23 \pm 2.75$ \\
& ResViT~\cite{dalmaz2022resvit} & $0.85 \pm 0.51$ ($P = 0.0005$) & $-0.27 \pm 2.28$ \\
& UNISYN (T1-only)~\cite{Zhang2023UnifiedMI} & $0.94 \pm 0.58$ ($P = 0.0097$) & $0.11 \pm 2.80$ \\
& Li-DiffNet~\cite{Zhang2026LiDiffNet} & $1.00 \pm 0.58$ ($P = 0.1233$) & $0.71 \pm 2.62$ \\
& UNISYN~\cite{Zhang2023UnifiedMI} & $1.04 \pm 0.60$ ($P = 0.1451$) & $1.16 \pm 2.94$ \\
& TriPF-Net (T1-only) & $1.04 \pm 0.60$ ($P = 0.1289$) & $0.41 \pm 2.77$ \\
& TriPF-Net & $\mathbf{1.08 \pm 0.58}$ ($P = 0.2699$) & $\mathbf{1.67 \pm 2.77}$ \\
\hline
\multirow{7}{*}{Dataset B}
& Pix2Pix~\cite{Isola_2017_CVPR} & $0.88 \pm 0.61$ ($P = 0.0173$) & $1.88 \pm 3.69$ \\
& ResViT~\cite{dalmaz2022resvit} & $0.84 \pm 0.58$ ($P = 0.0218$) & $0.83 \pm 3.62$ \\
& UNISYN (T1-only)~\cite{Zhang2023UnifiedMI} & $0.86 \pm 0.58$ ($P = 0.0131$) & $1.56 \pm 3.53$ \\
& Li-DiffNet~\cite{Zhang2026LiDiffNet} & $0.91 \pm 0.59$ ($P = 0.0253$) & $1.02 \pm 3.92$ \\
& UNISYN~\cite{Zhang2023UnifiedMI} & $0.93 \pm 0.62$ ($P = 0.0506$) & $2.51 \pm 3.63$ \\
& TriPF-Net (T1-only) & $0.88 \pm 0.59$ ($P = 0.0221$) & $2.19 \pm 3.93$ \\
& TriPF-Net & $\mathbf{0.94 \pm 0.60}$ ($P = 0.0586$) & $\mathbf{3.04 \pm 3.57}$ \\
\hline
\end{tabular}
\end{table*}

As shown in Table~\ref{tab:cnr_snr}, TriPF-Net achieved the best lesion-level performance among all synthesized methods on both datasets. On Dataset A, the mean CNR of real HBP images was 1.18, whereas TriPF-Net achieved the highest CNR among all synthesized methods ($1.08 \pm 0.58$), with no statistically significant difference from real HBP images ($P = 0.2699$). Visual examples in Fig.~\ref{fig7} further support this observation, where TriPF-Net generated HBP images with lesion contrast and SNR values that more closely matched those of the real images. This finding indicates that TriPF-Net can effectively preserve lesion contrast and approximate the enhancement characteristics of real clinical HBP images.

\begin{figure*}[!t]
  \centering
  \includegraphics[width=0.7\textwidth]{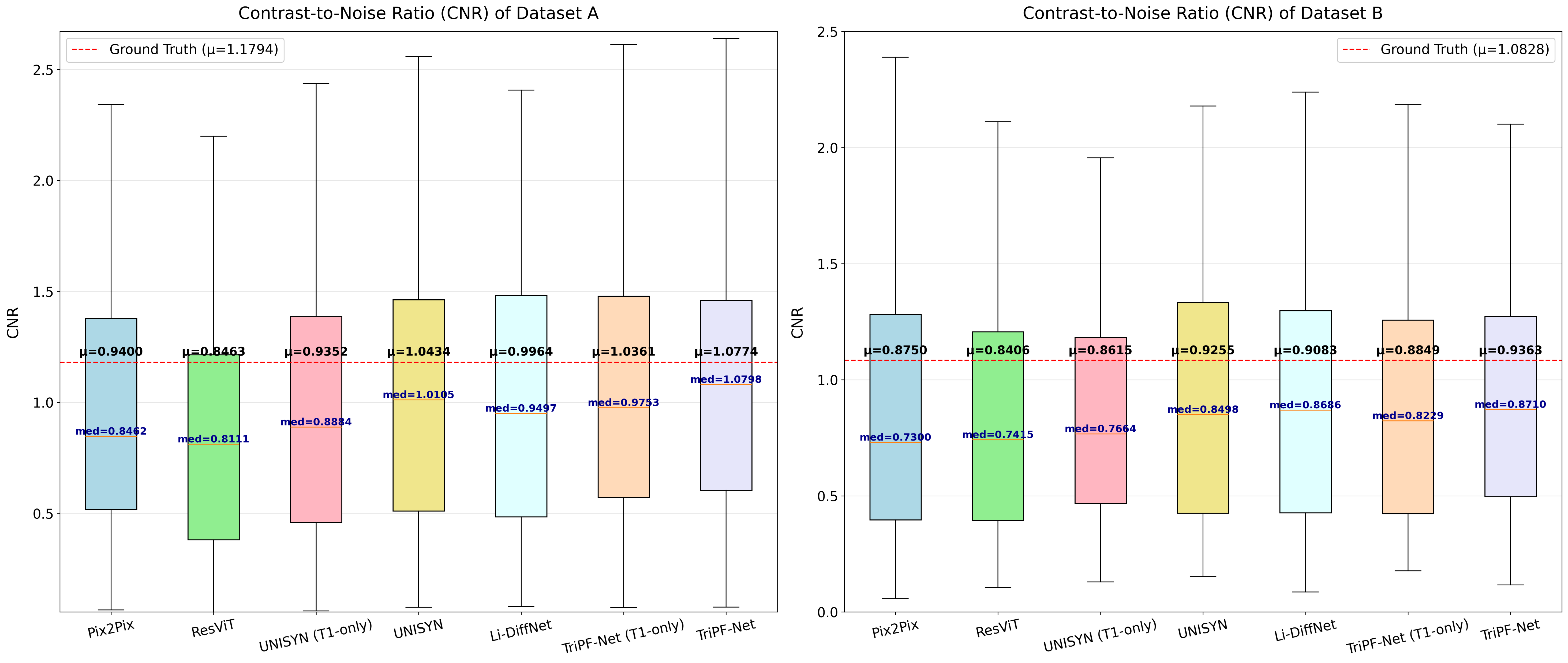}
  \caption{Boxplot comparison showing that TriPF-Net achieved the closest median CNR to real HBP images and a more compact distribution.}
  \label{fig6}
\end{figure*}

\begin{figure*}[!t]
  \centering
  \includegraphics[width=1.0\textwidth]{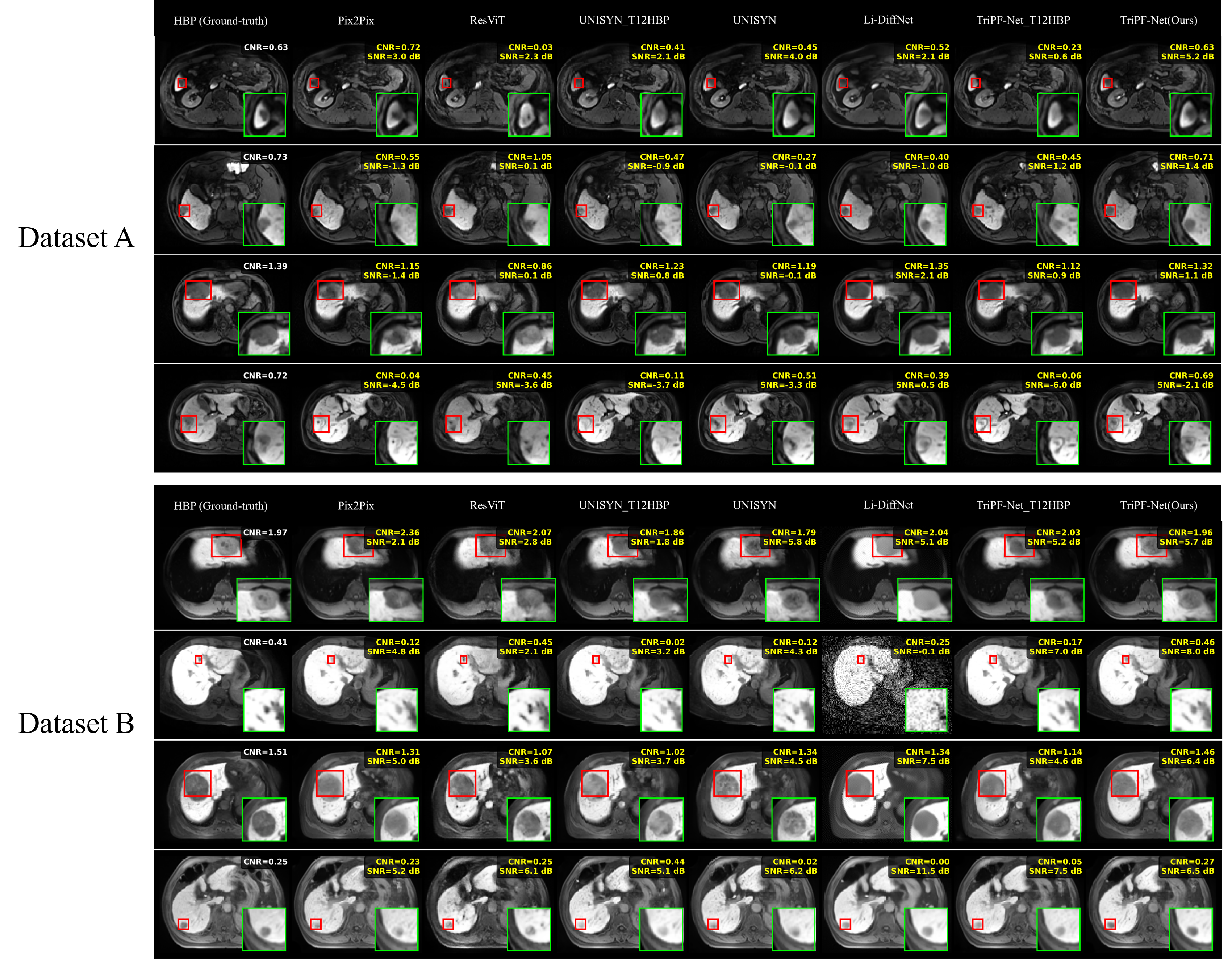}
  \caption{Representative synthetic HBP images produced by TriPF-Net and competing methods on Dataset A and Dataset B, with corresponding CNR and SNR values.}
  \label{fig7}
\end{figure*}

Although UNISYN, TriPF-Net (T1-only), and Li-DiffNet also showed no statistically significant difference from real images on Dataset A, the boxplot in Fig.~\ref{fig6} shows that TriPF-Net had the closest median CNR to the real value and a more compact distribution, suggesting better stability and lower variability in lesion depiction. In addition, TriPF-Net achieved the highest mean SNR among synthesized methods, indicating improved tumor-region signal characteristics. A similar trend was observed on Dataset B.

\subsection{Blinded Radiologist Evaluation}

\begin{table}[!t]
  \centering
  \caption{Blinded radiologist evaluation results on Datasets A and B (2 categories: real/synthetic). \emph{Real} and \emph{Synthetic} indicate radiologists' preferences.} 
  \label{tab:radiologist_evaluation_2cat}
  \begin{tabular}{lccc}
  \hline
  \textbf{Dataset} & \textbf{Radiologist} & \textbf{Real} & \textbf{Synthetic} \\
  \hline
  \multirow{2}{*}{Dataset A} & Radiologist 1 & 69.62\% & 30.38\% \\
                             & Radiologist 2 & 60.76\% & 39.24\% \\
  \hline
  \multirow{2}{*}{Dataset B} & Radiologist 1 & 70.59\% & 29.41\% \\
                             & Radiologist 2 & 72.55\% & 27.45\% \\
  \hline
  \end{tabular}
\end{table}

\begin{table*}[!t]
  \centering
  \caption{Blinded radiologist evaluation results on Datasets A and B (3 categories: real/synthetic/similar quality).}
  \label{tab:radiologist_evaluation_3cat}
  \begin{tabular}{lcccc}
  \hline
  \textbf{Dataset} & \textbf{Radiologist} & \textbf{Preferred Real} & \textbf{Preferred Synthetic} & \textbf{Similar Quality} \\
  \hline
  \multirow{2}{*}{Dataset A} & Radiologist 1 & 34.18\% & 7.59\% & 58.23\% \\
                             & Radiologist 2 & 20.25\% & 3.80\% & 75.95\% \\
  \hline
  \multirow{2}{*}{Dataset B} & Radiologist 1 & 54.90\% & 5.88\% & 39.22\% \\
                             & Radiologist 2 & 33.33\% & 5.88\% & 60.79\% \\
  \hline
  \end{tabular}
\end{table*}

To assess the clinical acceptability of the synthesized images, we conducted a blinded visual evaluation using a customized Python-based visual Turing test program, with each patient treated as an independent evaluation unit. The radiologists were blinded to image authenticity and based their decisions solely on subjective diagnostic preference. The program is publicly available at https://github.com/sxhjoker/TriPF-Net.

Table~\ref{tab:radiologist_evaluation_2cat} summarizes the results of the initial binary evaluation. On Dataset A, synthetic images were preferred in 30.38\% and 39.24\% of cases by Radiologist 1 and Radiologist 2, respectively; on Dataset B, the corresponding proportions were 29.41\% and 27.45\%. Although real images were preferred overall, the non-negligible proportion of cases favoring synthetic images should not be interpreted as evidence that synthesis outperforms true acquisition. Rather, it likely reflects the fact that real HBP images are not always of optimal quality—motion artifacts, respiratory misregistration, or suboptimal scan timing can degrade diagnostic clarity. In such cases, the artifact-free synthetic images may have appeared preferable, highlighting a potential role for TriPF-Net as a quality control reference: when a synthetic image is substantially preferred over the corresponding real HBP, it may signal that the real acquisition warrants closer inspection for quality issues.

To more accurately assess visual equivalence and reduce the bias introduced by forced binary selection, we conducted a revised blinded evaluation one month later by adding a third option, namely ``similar quality,'' in addition to ``prefer real image'' and ``prefer synthetic image.'' The results are shown in Table~\ref{tab:radiologist_evaluation_3cat}.

On Dataset A, Radiologist 1 rated real images as superior in 34.18\% of cases, synthetic images as superior in 7.59\%, and the two as similar in 58.23\%. Radiologist 2 rated real images as superior in 20.25\%, synthetic images as superior in 3.80\%, and the two as similar in 75.95\%. On Dataset B, Radiologist 1 reported 54.90\%, 5.88\%, and 39.22\%, respectively, whereas Radiologist 2 reported 33.33\%, 5.88\%, and 60.79\%, respectively.

Overall, although radiologists tended to prefer real images when they were of good quality, a considerable proportion of assessments rated the synthesized images as visually similar to real HBP images, with similarity rates reaching 75.95\% in one setting. Together with the binary evaluation results, these findings suggest that TriPF-Net generates HBP images with substantial visual realism, supporting its potential to streamline liver MRI workflows and serve as an automated quality control reference alongside delayed HBP acquisition.

\subsection{Tumor Segmentation for Quality Validation}

We further validated the fidelity of synthetic HBP images through a downstream tumor segmentation task using Attention U-Net~\cite{oktay2018attention}. To avoid data leakage, the generation training set (553 patients) was excluded; the remaining 237 patients (validation + test) were used in a three-fold cross-validation, with two folds (158) for training and one fold (79) for testing per round. A model trained on real HBP images under the same split served as the upper bound. All comparisons used the Dice similarity coefficient (DSC) and intersection over union (IoU).

As shown in Table~\ref{tab:seg}, the segmentation model trained on TriPF-Net outputs achieved a DSC of 70.41\% and an IoU of 56.25\%, approaching the real-HBP upper bound (DSC 71.61\%, IoU 57.48\%). A clearer gap was observed for UNISYN (DSC 70.22\%, IoU 56.06\%) and Pix2Pix (DSC 68.17\%, IoU 54.13\%). Notably, all DSC values are relatively modest because segmentation was not our primary objective: we used an off-the-shelf Attention U-Net without task-specific tuning or data curation, as the goal was to compare the relative fidelity of synthetic images rather than to maximize absolute segmentation performance. The results indicate that TriPF-Net-generated images preserve lesion-level anatomical information more faithfully, corroborating the pixel-level metrics reported above.

\begin{table}[!t]
  \centering
  \caption{Tumor segmentation performance on real HBP test images after training on real vs.\ synthetic HBP data, evaluated via three-fold cross-validation on Dataset A.}
  \label{tab:seg}
  \begin{tabular}{lcc}
  \hline
  \textbf{Training Data} & \textbf{DSC (\%)} & \textbf{IoU (\%)} \\
  \hline
  Real HBP  & 71.61 & 57.48 \\
  TriPF-Net (Ours) & 70.41 & 56.25 \\
  UNISYN~\cite{Zhang2023UnifiedMI}    & 70.22 & 56.06 \\
  Pix2Pix~\cite{Isola_2017_CVPR}   & 68.17 & 54.13 \\
  \hline
  \end{tabular}
  \end{table}

\subsection{Computational Cost}

We further compared the computational cost of TriPF-Net with that of the competing methods. Table~\ref{tab:cost} reports the number of model parameters at inference and training time, together with the computational cost per image. At inference, the three GAN-based methods, Pix2Pix~\cite{Isola_2017_CVPR}, UNISYN~\cite{Zhang2023UnifiedMI}, and TriPF-Net, do not use the discriminators, which causes the difference between the two parameter columns. Li-DiffNet~\cite{Zhang2026LiDiffNet}, by contrast, is diffusion-based and therefore shows no difference. TriPF-Net introduces only a marginal overhead over the UNISYN backbone on which it is built (+1.67M inference parameters and +0.5 GFLOPs per image), indicating that the ERGE, TPUM, and CEK modules are computationally lightweight. In contrast, Li-DiffNet requires approximately 14 times the computational cost of TriPF-Net (1848.31 vs.\ 133.07 GFLOPs per image) despite having far fewer parameters, reflecting the cost of iterative diffusion sampling.

\begin{table*}[!t]
  \centering
  \caption{Computational cost comparison of the competing methods.}
  \label{tab:cost}
  \setlength{\tabcolsep}{4pt}
  \begin{tabular}{lccc}
  \hline
  \textbf{Models} & \textbf{Params (inference)} & \textbf{Params (train)} & \textbf{GFLOPs/image} \\
  \hline
  Pix2Pix~\cite{Isola_2017_CVPR} & 43.10M & 45.86M & 53.16 \\
  UNISYN~\cite{Zhang2023UnifiedMI} & 84.10M & 95.16M & 132.57 \\
  Li-DiffNet~\cite{Zhang2026LiDiffNet} & 10.92M & 10.92M & 1848.31 \\
  TriPF-Net & 85.77M & 96.82M & 133.07 \\
  \hline
  \end{tabular}
\end{table*}

\section{Ablation Studies}
\label{ablation}

To further investigate the contribution of each component in TriPF-Net, we conducted ablation studies by systematically adding or modifying key elements of the framework. All experiments were performed on Dataset A using UNISYN~\cite{Zhang2023UnifiedMI} as the baseline model. We evaluated the effects of three components: the clinical feature encoding module, the Contrast Enhancement Kinetics (CEK) constraint, and the Gabor filtering layer.

As shown in Table~\ref{tab:Ablation}, incorporating the Gabor filtering layer yielded the most pronounced improvement, reducing MAE to 10.66 and increasing PSNR and SSIM to 23.36 and 0.759, respectively. The introduction of the CEK constraint also improved reconstruction quality, suggesting that contrast-enhancement kinetics provide useful physiological guidance for HBP synthesis. Since the CEK Module encodes precisely the enhancement kinetics quantified in Fig.~\ref{fig1}, this improvement directly measures the impact of overlooking versus modeling the characteristic enhancement dynamics of liver parenchyma and focal lesions. In contrast, adding only the clinical feature encoding module produced limited improvements, mainly in SSIM and FID.

Overall, the full TriPF-Net achieved the best balance across all metrics, supporting the complementary value of the proposed components.

\begin{table*}[!t]
  \centering
  \caption{Ablation study results on Dataset A.} 
  \label{tab:Ablation}
  \setlength{\tabcolsep}{4pt} 
  \begin{tabular}{lcccc}
  \hline
  \textbf{Models} & \textbf{MAE $\downarrow$} & \textbf{PSNR $\uparrow$} & \textbf{SSIM $\uparrow$} & \textbf{FID $\downarrow$} \\
  \hline
  Baseline & $11.31 \pm 4.02$ & $22.82 \pm 2.19$ & $0.747 \pm 0.083$ & 11.32 \\
  Baseline+Clinical & $11.45 \pm 4.07$ & $22.73 \pm 2.18$ & $0.751 \pm 0.080$ & 11.23 \\
  Baseline+CEK & $10.90 \pm 3.94$ & $23.11 \pm 2.24$ & $0.755 \pm 0.079$ & 11.07 \\
  Baseline+Gabor & $10.66 \pm 4.06$ & $23.36 \pm 2.27$ & $0.759 \pm 0.078$ & 10.58 \\
  Full TriPF-Net & $\mathbf{10.65 \pm 3.84}$ & $23.27 \pm 2.22$ & $\mathbf{0.761 \pm 0.077}$ & \textbf{10.49} \\
  \hline
  \end{tabular}
\end{table*}

\subsection{Sensitivity to the Loss Weights}

Finally, we examined the sensitivity of TriPF-Net to the two loss weights defined in Section~\ref{sec:loss}, $\lambda_{\mathrm{CEK}}$ and $\lambda_{\mathrm{adv}}$. Table~\ref{tab:lambda} reports the results obtained by varying one weight while keeping the other at its default value. Across all five configurations, MAE varies within 10.53--10.82 (about 2.7\% of the best value), PSNR within 23.22--23.45, and SSIM within 0.757--0.766, indicating that the performance of TriPF-Net is only weakly sensitive to both weights. For $\lambda_{\mathrm{CEK}}$, the default value of 50 is optimal or tied-optimal among the three settings examined. For $\lambda_{\mathrm{adv}}$, increasing the weight to 1.0 slightly improves the pixel-level metrics but degrades FID (10.91), whereas lowering it to 0.6 yields the best FID (10.37) at a small cost in the other metrics. The default setting of $\lambda_{\mathrm{CEK}}=50$ and $\lambda_{\mathrm{adv}}=0.8$ therefore provides the most balanced trade-off across the four metrics and was retained for all experiments reported in this study.

\begin{table*}[!t]
  \centering
  \caption{Sensitivity of TriPF-Net to the loss weights $\lambda_{\mathrm{CEK}}$ and $\lambda_{\mathrm{adv}}$ (Dataset A). The last row is the setting adopted in this study.}
  \label{tab:lambda}
  \setlength{\tabcolsep}{4pt}
  \begin{tabular}{lcccc}
  \hline
  \textbf{Setting} & \textbf{MAE $\downarrow$} & \textbf{PSNR $\uparrow$} & \textbf{SSIM $\uparrow$} & \textbf{FID $\downarrow$} \\
  \hline
  $\lambda_{\mathrm{CEK}}$=40, $\lambda_{\mathrm{adv}}$=0.8 & $10.82 \pm 4.00$ & $23.22 \pm 2.25$ & $0.757 \pm 0.078$ & 10.66 \\
  $\lambda_{\mathrm{CEK}}$=60, $\lambda_{\mathrm{adv}}$=0.8 & $10.71 \pm 3.98$ & $23.27 \pm 2.29$ & $0.757 \pm 0.079$ & 10.73 \\
  $\lambda_{\mathrm{CEK}}$=50, $\lambda_{\mathrm{adv}}$=0.6 & $10.75 \pm 4.05$ & $23.23 \pm 2.27$ & $0.758 \pm 0.079$ & 10.37 \\
  $\lambda_{\mathrm{CEK}}$=50, $\lambda_{\mathrm{adv}}$=1.0 & $10.53 \pm 4.03$ & $23.45 \pm 2.31$ & $0.766 \pm 0.075$ & 10.91 \\
  $\lambda_{\mathrm{CEK}}$=50, $\lambda_{\mathrm{adv}}$=0.8 (default) & $10.65 \pm 3.84$ & $23.27 \pm 2.22$ & $0.761 \pm 0.077$ & 10.49 \\
  \hline
  \end{tabular}
\end{table*}

\section{Conclusions}
\label{conclusions}
In this study, we developed TriPF-Net, a flexible and physiologically informed framework for synthesizing hepatobiliary-phase (HBP) MRI in patients with hepatocellular carcinoma. Moving beyond conventional static fusion methods, TriPF-Net explicitly captures the sequential enhancement dynamics of liver parenchyma and focal lesions while maintaining high robustness against the stochastic absence of arterial- or venous-phase inputs. By uniquely integrating multi-phase imaging with patient-specific clinical biomarkers, the proposed model generates HBP images that are both visually realistic and physiologically plausible.
Comprehensive validation across two clinical centers demonstrated that TriPF-Net consistently outperforms state-of-the-art methods in both global reconstruction and lesion-level contrast. Notably, the synthesized images achieved contrast-to-noise ratios (CNR) comparable to real HBP images and gained strong endorsement in blinded radiologist evaluations. These findings suggest that TriPF-Net provides a reliable solution for streamlining liver MRI workflows. Beyond reducing dependence on delayed HBP acquisitions, the synthetic images may serve as an automated quality control reference: when a synthetic HBP is substantially preferred over its real counterpart, it can alert clinicians to potential degradation in the acquired image that warrants closer inspection.

\section{Ethics Statement and Patient Consent}
Dataset A was collected from The First Affiliated Hospital (Southwest Hospital) of Army Medical University, and Dataset B was collected from The First Affiliated Hospital of Nanjing Medical University.
The study protocol was approved by the Ethics Committee of The First Affiliated Hospital of Army Medical University (Reference Number: KY2024126, KY2024209).
All cases were analyzed anonymously, with no patient-identifiable information involved.
The source code of TriPF-Net (including the ERGE, TPUM, and CEK modules), the training pipeline, the data preprocessing scripts, the trained weights, and the visual Turing test program are publicly available at https://github.com/sxhjoker/TriPF-Net. The datasets used in this study are not publicly available due to institutional privacy restrictions but are available from the corresponding author upon reasonable request.

\section{Acknowledgements}
This work was supported by 
the National Natural Science Foundation of China (No. 62601822),
the Special Project for Enhancing Technological Innovation Capability of Army Medical University (No. 2023XQN31),
the Chongqing Postdoctoral Special Funding Project (No. 2024CQBSHTB3009),
the Joint project of Chongqing Health Commission and Science and Technology Bureau (No. 2025MSXM016), 
and the Independent Project of Yu-Yue Pathology Scientific Research Center (No. YYKYXM202303AZ-205).

The manuscript was refined with the assistance of DeepSeek for language polishing and grammatical improvements.
 \bibliographystyle{elsarticle-num}
 \bibliography{cas-refs}

\end{document}